\documentclass[12pt,oneside,english,american]{book}

\usepackage[T1]{fontenc}
\usepackage[latin9]{inputenc}
\usepackage[letterpaper]{geometry}
\usepackage{fancyhdr}
\usepackage{array}
\usepackage{booktabs}
\usepackage{multirow}
\usepackage{amsmath}
\usepackage{amssymb}
\usepackage{mathdots}
\usepackage{setspace}
\PassOptionsToPackage{version=3}{mhchem}
\usepackage{mhchem}
\usepackage{esint}
\makeatletter

\providecommand{\tabularnewline}{\\}

\date{}
\usepackage{times}

\usepackage{amsthm}\usepackage{graphics}

\numberwithin{equation}{chapter}
\usepackage{babel}
\usepackage{titlesec} 

\titleformat{\chapter}{\centering\LARGE\bfseries}{\chaptertitlename\ \thechapter}{0.7em}{}
\titlespacing*{\chapter}{0pt}{0pt}{50pt}

\titleformat{\section}{\large\bfseries}{\thesection}{0.7em}{}
\titlespacing*{\section}{0pt}{12pt}{0pt}

\usepackage{indentfirst}
\usepackage{mathtools}

\makeatother

\usepackage{babel}
\begin{document}
\begin{titlepage}

\title{{\Large{}An MCMC Algorithm for Estimating the Reduced RUM}}
\begin{spacing}{0.050000000000000003}

\author{\noindent Meng-ta Chung\textsuperscript{1} and Matthew S. Johnson\textsuperscript{2}\vspace{+20pt}\\\textsuperscript{1}
Department of Human Development, Columbia University\\\textsuperscript{2}
Office of Medical Research, St. Joseph's Hospital \\}
\end{spacing}

\maketitle
\thispagestyle{empty} \end{titlepage}
\noindent \begin{center}
\textbf{\Large{}Abstract}
\par\end{center}{\Large \par}

\noindent The RRUM is a model that is frequently seen in language
assessment studies. The objective of this research is to advance an
MCMC algorithm for the Bayesian RRUM. The algorithm starts with estimating
correlated attributes. Using a saturated model and a binary decimal
conversion, the algorithm transforms possible attribute patterns to
a Multinomial distribution. Along with the likelihood of an attribute
pattern, a Dirichlet distribution is used as the prior to sample from
the posterior. The Dirichlet distribution is constructed using Gamma
distributions. Correlated attributes of examinees are generated using
the inverse transform sampling. Model parameters are estimated using
the Metropolis within Gibbs sampler sequentially. Two simulation studies
are conducted to evaluate the performance of the algorithm. The first
simulation uses a complete and balanced Q-matrix that measures 5 attributes.
Comprised of 28 items and 9 attributes, the Q-matrix for the second
simulation is incomplete and imbalanced. The empirical study uses
the ECPE data obtained from the CDM R package. Parameter estimates
from the MCMC algorithm and from the CDM R package are presented and
compared. The algorithm developed in this research is implemented
in R.\textbf{\vspace{10pt}}

\noindent \textbf{Keywords} 

\noindent CDM, RUM, RRUM, Q-matrix, Bayesian, MCMC

\newpage{}
\noindent \begin{center}
\textbf{\Large{}Introduction}
\par\end{center}{\Large \par}

\noindent Cognitive diagnostic assessment (CDA) is a framework that
aims to evaluate whether an examinee has mastered a particular cognitive
process called \textit{attribute} (Leighton \& Gierl, 2007). In CDA,
exam items are each associated with attributes that are required for
mastery. Using examinees' attribute states, CDA provides effective
information for examinees to improve their learning and for educators
to adjust their teaching. Studies have demonstrated that CDA is a
valid application for providing useful diagnostic feedback in language
assessment (e.g., Jang, 2009; Jang et al., 2013; Kim, 2011; Kim, 2014;
Li \& Suen, 2012; Richards, 2008). Concerning the current thinking
and future directions of CDA\textit{, Language Testing} publishes
a special issue (Volume 32, Issue 3, July 2015) that integrates insights
from experts in the field of language assessment.

In recent years, a few cognitive diagnosis models (CDMs) have been
developed, including the deterministic input, noisy-and gate (DINA)
model (Junker \& Sijtsma, 2001), the noisy input, deterministic-and
gate (NIDA) model (Maris, 1999), and the reparameterized unified model
(RUM) (Hartz, 2002; Hartz, Roussos, \& Stout, 2002). All these models
use the Q-matrix (Tatsuoka, 1983) to measure attribute states of examinees.
Suppose there are $I$ examinees taking the exam that measures $K$
attributes. A binary matrix $\boldsymbol{A}_{I\times K}=(\alpha_{ik})_{I\times K}$
reveals the connection between examinees and attributes. If examinee
$i$ does not mater attribute $k,$ then $\alpha_{ik}=0$; if examinee
$i$ masters attribute $k$, then $\alpha_{ik}=1$. 

In order to evaluate examinees with respect to their levels of competence
of each attribute in an exam, the Q-matrix (Tatsuoka, 1983) is used
to partition exam items into attributes. The Q-matrix is a binary
matrix that shows the relationship between exam items and attributes.
Given an exam with $J$ items that measure $K$ attributes, the Q-matrix
is represented as a $J$ by $K$ matrix, $\boldsymbol{Q}_{J\times K}=(q_{jk})_{J\times K}$.
In a Q-matrix, if attribute $k$ is required by item $j,$ then $q_{jk}=1$.
If attribute $k$ is not required by item $j$, then $q_{jk}=0$. 

Among all of the CDMs, the RUM is frequently seen in language assessment
research. Extending the NIDA model, Maris (1999) proposed a model
that attempts to estimate the slip and guess parameters for different
items. That is, the the slip and guess parameters have subscripts
for both items and attributes. To improve this model, Dibello, Stout,
and Roussos (1995) advances the unified model that incorporates a
unidimensional ability parameter. However, these two models are not
statistically identifiable. Hartz (2002) reparameterizes the unified
model so that the parameters of the model can be identified while
retaining their interpretability. As is expected, this reparameterized
unified model is a more complicated conjunctive CDMs (Roussos, Templin,
\& Hensen, 2007). The RUM defines the probability of a correct response
to an item as \begin{equation} \pi_{j}^{*}=\prod\limits _{k=1}^{K}(1-s_{jk})^{q_{jk}}, \tag{1}\end{equation}and
the penalty for each attribute no possessed as \begin{equation} r_{jk}^{*}=g_{jk}/1-s_{jk}. \tag{2}\end{equation}

$\pi_{j}^{*}$ is the probability that an examinee, having acquired
all the attributes required for item $j$, will correctly apply these
attributes in solving the item. That is, $\pi^{*}$ is interpreted
as an item difficulty parameter. $r_{jk}^{*}$ is used to define the
penalty of not mastering the $k^{th}$ attribute. Under this view,
$r_{jk}^{*}$ can be seen as an indicator of the diagnostic capacity
of item $j$ for attribute $k$. Also from the perspective of monotonicity,
$1-s_{jk}$ should be greater than $g_{jk}$. Explicitly, $r_{jk}^{*}$
should be constrained to the interval $(0,1)$.

Incorporating a general ability measure, $P_{c_{j}}(\theta_{i})$,
the probability of a correct response in the RUM can be written as
\[
P(X_{ij}=1|\boldsymbol{\alpha},r^{*},\pi^{*},\theta)=\pi_{j}^{*}\prod\limits _{k=1}^{K}(r_{jk}^{*^{(1-\alpha_{ik})}})^{q_{jk}}P_{c_{j}}(\theta_{i}).
\]
$P_{c_{j}}(\theta_{i})$ is the item characteristic curve in the Rasch
model, where $c_{j}$ is the difficulty parameter and $\theta_{i}$
is the general measure of an examinee's knowledge not specified by
the Q-matrix.

The RUM has larger flexibility than other CDMs in modeling the probability
of correct item response for different attribute patterns. This flexibility,
however, is achieved at the cost of introducing a significant degree
of complexity into the estimation process. Assuming that the Q-matrix
completely specifies the attributes required by the exam items, Hartz
(2002) further suggests a reduced version of the RUM (RRUM) that sets
$P_{c_{j}}(\theta_{i})=1$. The parameters of the RRUM retain the
model identifiable and allow the probabilities of slipping and guessing
to vary across items. The IRF of the RRUM is therefore reduced to
\begin{equation} P(X_{ij}=1|\boldsymbol{\alpha},r^{*},\pi^{*})=\pi_{j}^{*}\prod\limits _{k=1}^{K}(r_{jk}^{*^{(1-\alpha_{ik})}})^{q_{jk}}.  \tag{3}\end{equation}Based
on the assumptions of local independence and independence among examinees,
the joint likelihood function for all responses in the RRUM is 
\begin{multline*}
P(X_{ij}=x_{ij},\forall i,j|\boldsymbol{\alpha},r^{*},\pi^{*})\\
=\prod\limits _{i=1}^{I}\prod\limits _{j=1}^{J}\Biggl(\pi_{j}^{*}\prod\limits _{k=1}^{K}r_{jk}^{*^{(1-\alpha_{jk})q_{jk}}}\Biggr)^{x_{ij}}\Biggl(1-\pi_{j}^{*}\prod\limits _{k=1}^{K}r_{jk}^{*^{(1-\alpha_{jk})q_{jk}}}\Biggr)^{1-x_{ij}}.
\end{multline*}

The RRUM is a simplified yet practical model that has received considerable
attention among psychometricians and educators (e.g., Chiu \& Köhn,
2016; Feng, Habing, \& Huebner, 2014; Henson \& Templin, 2007; Jang,
2009; Jang et al., 2013; Kim, 2011; Kim, 2014; Templin, 2004; Templin
et al., 2004; Templin \& Douglas, 2004; Zhang, 2013). Nevertheless,
the RRUM remains more complex than other CDMs. Due to its complexity,
the RRUM have been mostly estimated in a Bayesian framework. Hartz
(2002) uses a Bayesian method to estimate the RRUM, and Hartz, Roussos,
\& Stout (2002) develops the patented \textit{Arpeggio} program, which
is commonly applied to analyze the data in language assessment studies. 

This research proposes a different MCMC algorithm for estimating the
Bayesian RRUM, with the hope of reducing the complexity of computation.
Specifically, a saturated model using the inverse transform sampling
is used to estimate correlated attributes, and the Metropolis with
Gibbs sampling is adopted to estimate the $\boldsymbol{\pi}^{*}$
and $\boldsymbol{r}^{*}$ parameters. The proposed algorithm, as well
as a way to simulate data, are implemented in R (R Development Core
Team, 2017). With the algorithm, it is readily flexible for researchers
and practitioners to code using any programming languages.
\noindent \begin{center}
\textbf{\Large{}Proposed MCMC Algorithm}
\par\end{center}{\Large \par}

\noindent The setting for the estimation is comprised of responses
from $I$ examinees to $J$ items that measure $K$ attributes. Given
a $J$ by $K$ $Q$-matrix, the following steps perform sequentially
at iteration $t$, $t=1,\ldots,T$.

\noindent \textbf{Step 1: Binary Decimal Conversion}

\noindent With $K$ attributes, there are a total of $2^{K}$ \textit{possible}
attribute patterns for examinee $i$. Let $2^{K}=M$, and let the
matrix $\boldsymbol{x}_{M\times K}=(x_{mk})_{M\times K}$ be the matrix
of \textit{possible} attribute patterns. Each of the $M$ rows in
$\boldsymbol{x}$ represents a possible attribute pattern, which is
converted to a decimal number by $(b_{n}b_{n-1}\cdots b_{0})_{2}=b_{n}(2)^{n}+b_{n-1}(2)^{n-1}+\cdots+b_{0}(2)^{0},$
where $(b_{n}b_{n-1}\cdots b_{0})_{2}$ denotes a binary number.

After the conversion, these $M$ possible attribute patterns become
a Multinomial distribution. To estimate correlated attributes, a saturated
Multinomial model is used that assumes no restrictions on the probabilities
of the attribute patterns (see Maris, 1999). Assuming a Dirichlet
prior $\boldsymbol{\theta}$, the hierarchical model for estimating
attributes is\textbf{\vspace{-5pt}} 
\begin{align*}
\boldsymbol{x}|\boldsymbol{\theta} & \sim M\!ultinomial(M,\boldsymbol{\theta}),\\
\boldsymbol{\theta} & \sim Dirichlet(a_{1},a_{2},\ldots,a_{M}).
\end{align*}

\noindent \textbf{Step2: Updating Probability of Attribute Pattern}

\noindent Let $\boldsymbol{y}$ and $\boldsymbol{q}$ be the data
and the Q-matrix. The full conditional posterior distribution is $p(\boldsymbol{x}|\boldsymbol{y},\boldsymbol{\pi}^{*},\boldsymbol{r}^{*},\boldsymbol{q})\propto p(\boldsymbol{y}|\boldsymbol{x},\boldsymbol{\pi}^{*},\boldsymbol{r}^{*},\boldsymbol{q})p(\boldsymbol{x}|\boldsymbol{\theta})p(\boldsymbol{\theta})$.
As the conjugate prior for a Multinomial distribution is also a Dirichlet
distribution, $p(\boldsymbol{x}|\boldsymbol{\theta})p(\boldsymbol{\theta})$
is a Dirichlet distribution. Therefore, use $Dirichlet(1,1,\ldots,1)$
as the prior, and the conditional posterior is distributed as $\mathit{Dirichlet}(1+y_{1},1+y_{2},\ldots,1+y_{M}),$
where $y_{\ell}$ $(\ell=1,\ldots,M)$ is the number examinees possessing
the $\ell^{th}$ attribute pattern. As no function in base R can be
used to sample from Dirichlet distribution, Gamma distributions are
used to construct the Dirichlet distribution. In step 2, suppose that
$w_{1},\ldots,w_{M}$ are distributed as $\mathit{Gamma}(a_{1},1),\ldots,\mathit{Gamma}(a_{M},1)$,
and that $\tau=w_{1}+\cdots+w_{M}$, then $(w_{1}/\tau,w_{2}/\tau,\ldots,w_{M}/\tau)$
is distributed as $\mathit{Dirichlet}(a_{1},a_{2},\ldots,a_{M})$. 

For each of the $M$ possible attribute patterns, step 2 calculates
the total number of examinees $(y_{1},y_{2},\ldots,y_{M})$ falling
into an attribute pattern, and then samples from $\mathit{Gamma}(1+y_{1},1)=w'_{1},Gamma(1+y_{2},1)=w'_{2},\ldots,Gamma(1+y_{M},1)=w'_{M}$.
Let $\tau'=w'_{1}+w'_{2}+\cdots+w'_{M}$, and we can get $p(\boldsymbol{x}|\boldsymbol{\theta})p(\boldsymbol{\theta})=(w'_{1}/\tau',w'_{2}/\tau',\ldots,w'_{M}/\tau')$.
Along with the likelihood of each possible attribute pattern, which
is $p(\boldsymbol{y}|\boldsymbol{x},\boldsymbol{\pi}^{*},\boldsymbol{r}^{*},\boldsymbol{q})$,
step 2 obtains the full conditional posterior.

\noindent \textbf{Step 3: Updating Attribute}

\noindent The full conditional posterior distribution is sampled using
the discrete version of inverse transform sampling. Let the posterior
$(p_{1},p_{2},\text{\dots},p_{M})$ be the PMF of the $M$ possible
attribute patterns. The CDF is computed by adding up the probabilities
for the $M$ points of the distribution. To sample from this discrete
distribution, we partition $(0,1)$ into $M$ subintervals $(0,p_{1})$,
$(p_{1},p_{1}+p_{2})$, \ldots{}, $(\sum\limits _{m=0}^{M}p_{m-1},\sum\limits _{m=0}^{M}p_{m})$,
and then generate a value $u$ from $\mathit{Uniform}(0,1)$. 

Updating the attribute state of examinee $i$ is achieved by checking
which subinterval the value $u$ falls into. This subinterval number
(a decimal number) is then converted to its corresponding binary number
(see step 1) that represents the attribute state of examinee $i$.
After step 3 is applied to each examinee, attribute states for all
examinees, denoted as $\boldsymbol{\alpha}$, are obtained for iteration
$t$. 

\noindent \textbf{Step 4: Updating $\boldsymbol{r}^{*}$ and $\boldsymbol{\pi}^{*}$
Parameters}

\noindent A Metropolis within Gibbs algorithm is used to sample $\boldsymbol{\pi}^{*}$
and $\boldsymbol{r}^{*}$. The non-informative $Beta(1,1)$ prior
is applied in updating $\boldsymbol{r}^{*}$ and $\boldsymbol{\pi}^{*}$.
Candidate values for $\boldsymbol{r}^{*}$ is sampled from $\mathit{Uniform}(\boldsymbol{r}^{*^{(t-1)}}-\delta,\boldsymbol{r}^{*^{(t-1)}}+\delta)$.
It should be noted that candidate values for $\boldsymbol{r}^{*}$
are restricted to the interval $(0,1)$, and that $\delta$ is adjusted
so that the acceptance rate is between 25\% and 40\% (see Gilks et
al., 1996). The updated $\boldsymbol{\alpha}$ from step 3 is carried
to step 4. As $\boldsymbol{\pi}^{*}$ and $\boldsymbol{r}^{*}$ are
assumed to be independent of each other, $p(\boldsymbol{\pi}^{*},\boldsymbol{r}^{*})=p(\boldsymbol{\pi}^{*})p(\boldsymbol{r}^{*})$. 

In updating $\boldsymbol{r}^{*}$ at iteration $t$, the acceptance
probability $\varphi_{r}$ for the candidate value $\boldsymbol{r}^{*^{(*)}}$
is calculated by \vspace{10pt} 
\begin{align*}
\varphi_{r} & =\dfrac{p(\boldsymbol{y}|\boldsymbol{\alpha}^{(t)},\boldsymbol{r}^{*^{(*)}},\boldsymbol{\pi}^{*^{(t-1)}},\boldsymbol{q})p(\boldsymbol{r}^{*^{(*)}})}{p(\boldsymbol{y}|\boldsymbol{\alpha}^{(t)},\boldsymbol{r}^{*^{(t-1)}},\boldsymbol{\pi}^{*^{(t-1)}},\boldsymbol{q})p(\boldsymbol{r}^{*^{(t-1)}})},
\end{align*}
and $\boldsymbol{r}^{*^{(t)}}$ is then set by
\[
\boldsymbol{r}^{*^{(t)}}=\begin{cases}
\boldsymbol{r}^{*^{(*)}} & \mathrm{\textrm{with\ probability\ min}}(1,\varphi)\\
\boldsymbol{r}^{*^{(t-1)}} & \textrm{otherwise}
\end{cases}.
\]
With the obtained $\boldsymbol{r}^{*^{(t)}}$, the acceptance probability
for updating $\boldsymbol{\pi}^{*}$ is 
\begin{align*}
\varphi_{\pi} & =\dfrac{p(\boldsymbol{y}|\boldsymbol{\alpha}^{(t)},\boldsymbol{r}^{*},\boldsymbol{\pi}^{*^{(t)}},\boldsymbol{q})p(\boldsymbol{\pi}^{*^{(t)}})}{p(\boldsymbol{y}|\boldsymbol{\alpha}^{(t)},\boldsymbol{r}^{*},\boldsymbol{\pi}^{*^{(t-1)}},\boldsymbol{q})p(\boldsymbol{\pi}^{*^{(t-1)}})},
\end{align*}
and $\boldsymbol{\pi}^{*^{(t)}}$ is decided by
\[
\boldsymbol{\pi}^{*^{(t)}}=\begin{cases}
\boldsymbol{\pi}^{*^{(*)}} & \mathrm{\textrm{with\ probability\ min}}(1,\varphi)\\
\boldsymbol{\pi}^{*^{(t-1)}} & \textrm{otherwise}
\end{cases}.
\]
\noindent \begin{center}
\textbf{\Large{}Simulation Study}
\par\end{center}{\Large \par}

\noindent \textbf{Procedure for Simulating Data}

\noindent To investigate the effectiveness of the proposed MCMC algorithm,
simulation studies are conducted to see how well the true attribute
states could be recovered. Simulated data sets are generated using
the following procedure. 

\selectlanguage{english}%
\begin{table}[h]
\begin{onehalfspace}
\caption{\foreignlanguage{american}{\label{tab:3.1}Q-matrix for Simulation I}}

\end{onehalfspace}

\selectlanguage{american}%
\label{tab:tab2-1-2-1} \centering\foreignlanguage{english}{}%
\begin{tabular*}{15cm}{@{\extracolsep{\fill}}ccccccccccccccc}
\toprule 
\multirow{2}{*}{{\small{}Item }} &  & \multicolumn{5}{c}{{\small{}Attribute}} &  & \multirow{2}{*}{{\small{}Item}} &  & \multicolumn{5}{c}{{\small{}Attribute}}\tabularnewline
\cmidrule{3-7} \cmidrule{11-15} 
 &  & {\small{}1 } & {\small{}2} & {\small{}3} & {\small{}4} & {\small{}5} &  &  &  & {\small{}1} & {\small{}2} & {\small{}3} & {\small{}4} & {\small{}5}\tabularnewline
\cmidrule{1-1} \cmidrule{3-7} \cmidrule{9-9} \cmidrule{11-15} 
{\small{}1} &  & {\small{}1} & {\small{}0} & {\small{}0} & {\small{}0} & {\small{}0} &  & {\small{}16} &  & {\small{}0} & {\small{}1} & {\small{}0} & {\small{}1} & {\small{}0}\tabularnewline
{\small{}2} &  & {\small{}0} & {\small{}1} & {\small{}0} & {\small{}0} & {\small{}0} &  & {\small{}17} &  & {\small{}0} & {\small{}1} & {\small{}0} & {\small{}0} & {\small{}1}\tabularnewline
{\small{}3} &  & {\small{}0} & {\small{}0} & {\small{}1} & {\small{}0} & {\small{}0} &  & {\small{}18} &  & {\small{}0} & {\small{}0} & {\small{}1} & {\small{}1} & {\small{}0}\tabularnewline
{\small{}4} &  & {\small{}0} & {\small{}0} & {\small{}0} & {\small{}1} & {\small{}0} &  & {\small{}19} &  & {\small{}0} & {\small{}0} & {\small{}1} & {\small{}0} & {\small{}1}\tabularnewline
{\small{}5} &  & {\small{}0} & {\small{}0} & {\small{}0} & {\small{}0} & {\small{}1} &  & {\small{}20} &  & {\small{}0} & {\small{}0} & {\small{}0} & {\small{}1} & {\small{}1}\tabularnewline
{\small{}6} &  & {\small{}1} & {\small{}0} & {\small{}0} & {\small{}0} & {\small{}0} &  & {\small{}21} &  & {\small{}1} & {\small{}1} & {\small{}1} & {\small{}0} & {\small{}0}\tabularnewline
{\small{}7} &  & {\small{}0} & {\small{}1} & {\small{}0} & {\small{}0} & {\small{}0} &  & {\small{}22} &  & {\small{}1} & {\small{}1} & {\small{}0} & {\small{}1} & {\small{}0}\tabularnewline
{\small{}8} &  & {\small{}0} & {\small{}0} & {\small{}1} & {\small{}0} & {\small{}0} &  & {\small{}23} &  & {\small{}1} & {\small{}1} & {\small{}0} & {\small{}0} & {\small{}1}\tabularnewline
{\small{}9} &  & {\small{}0} & {\small{}0} & {\small{}0} & {\small{}1} & {\small{}0} &  & {\small{}24} &  & {\small{}1} & {\small{}0} & {\small{}1} & {\small{}1} & {\small{}0}\tabularnewline
{\small{}10} &  & {\small{}0} & {\small{}0} & {\small{}0} & {\small{}0} & {\small{}1} &  & {\small{}25} &  & {\small{}1} & {\small{}0} & {\small{}1} & {\small{}0} & {\small{}1}\tabularnewline
{\small{}11} &  & {\small{}1} & {\small{}1} & {\small{}0} & {\small{}0} & {\small{}0} &  & {\small{}26} &  & {\small{}1} & {\small{}0} & {\small{}0} & {\small{}1} & {\small{}1}\tabularnewline
{\small{}12} &  & {\small{}1} & {\small{}0} & {\small{}1} & {\small{}0} & {\small{}0} &  & {\small{}27} &  & {\small{}0} & {\small{}1} & {\small{}1} & {\small{}1} & {\small{}0}\tabularnewline
{\small{}13} &  & {\small{}1} & {\small{}0} & {\small{}0} & {\small{}1} & {\small{}0} &  & {\small{}28} &  & {\small{}0} & {\small{}1} & {\small{}1} & {\small{}0} & {\small{}1}\tabularnewline
{\small{}14} &  & {\small{}1} & {\small{}0} & {\small{}0} & {\small{}0} & {\small{}1} &  & {\small{}29} &  & {\small{}0} & {\small{}1} & {\small{}0} & {\small{}1} & {\small{}1}\tabularnewline
{\small{}15} &  & {\small{}0} & {\small{}1} & {\small{}1} & {\small{}0} & {\small{}0} &  & {\small{}30} &  & {\small{}0} & {\small{}0} & {\small{}1} & {\small{}1} & {\small{}1}\tabularnewline
\bottomrule
\end{tabular*}\selectlanguage{english}%
\end{table}

\begin{table}[h]
\begin{onehalfspace}
\caption{\foreignlanguage{american}{\label{tab:3.1-1}Q-matrix for Simulation II}}

\end{onehalfspace}

\selectlanguage{american}%
\label{tab:tab2-1-2-1-1} \centering\foreignlanguage{english}{}%
\begin{tabular*}{15cm}{@{\extracolsep{\fill}}ccccccccccccccccccccccc}
\hline 
\multirow{2}{*}{{\small{}Item }} &  & \multicolumn{9}{c}{{\small{}Attribute}} &  & \multirow{2}{*}{{\small{}Item}} &  & \multicolumn{9}{c}{{\small{}Attribute}}\tabularnewline
\cline{3-11} \cline{15-23} 
 &  & {\small{}1 } & {\small{}2} & {\small{}3} & {\small{}4} & {\small{}5} & {\small{}6} & {\small{}7} & {\small{}8} & {\small{}9} &  &  &  & {\small{}1} & {\small{}2} & {\small{}3} & {\small{}4} & {\small{}5} & {\small{}6} & {\small{}7} & {\small{}8} & {\small{}9}\tabularnewline
\cline{1-1} \cline{3-11} \cline{13-13} \cline{15-23} 
{\small{}1} &  & {\small{}1} & {\small{}0} & {\small{}1} & {\small{}0} & {\small{}0} & {\small{}0} & {\small{}0} & {\small{}0} & {\small{}0} &  & {\small{}20} &  & {\small{}0} & {\small{}0} & {\small{}0} & {\small{}1} & {\small{}0} & {\small{}0} & {\small{}0} & {\small{}1} & {\small{}0}\tabularnewline
{\small{}2} &  & {\small{}0} & {\small{}0} & {\small{}1} & {\small{}0} & {\small{}0} & {\small{}1} & {\small{}0} & {\small{}0} & {\small{}0} &  & {\small{}21} &  & {\small{}0} & {\small{}1} & {\small{}0} & {\small{}0} & {\small{}0} & {\small{}0} & {\small{}0} & {\small{}0} & {\small{}0}\tabularnewline
{\small{}3} &  & {\small{}0} & {\small{}1} & {\small{}1} & {\small{}0} & {\small{}0} & {\small{}0} & {\small{}0} & {\small{}0} & {\small{}0} &  & {\small{}22} &  & {\small{}0} & {\small{}0} & {\small{}1} & {\small{}1} & {\small{}0} & {\small{}0} & {\small{}1} & {\small{}0} & {\small{}0}\tabularnewline
{\small{}4} &  & {\small{}1} & {\small{}0} & {\small{}1} & {\small{}0} & {\small{}1} & {\small{}0} & {\small{}0} & {\small{}0} & {\small{}0} &  & {\small{}23} &  & {\small{}0} & {\small{}0} & {\small{}0} & {\small{}0} & {\small{}0} & {\small{}1} & {\small{}0} & {\small{}0} & {\small{}0}\tabularnewline
{\small{}5} &  & {\small{}0} & {\small{}0} & {\small{}0} & {\small{}0} & {\small{}1} & {\small{}0} & {\small{}1} & {\small{}1} & {\small{}0} &  & {\small{}24} &  & {\small{}0} & {\small{}0} & {\small{}1} & {\small{}1} & {\small{}0} & {\small{}0} & {\small{}0} & {\small{}0} & {\small{}0}\tabularnewline
{\small{}6} &  & {\small{}0} & {\small{}0} & {\small{}0} & {\small{}0} & {\small{}1} & {\small{}0} & {\small{}0} & {\small{}0} & {\small{}0} &  & {\small{}25} &  & {\small{}0} & {\small{}0} & {\small{}0} & {\small{}1} & {\small{}0} & {\small{}0} & {\small{}0} & {\small{}1} & {\small{}0}\tabularnewline
{\small{}7} &  & {\small{}0} & {\small{}1} & {\small{}0} & {\small{}0} & {\small{}0} & {\small{}1} & {\small{}1} & {\small{}0} & {\small{}0} &  & {\small{}26} &  & {\small{}0} & {\small{}0} & {\small{}1} & {\small{}0} & {\small{}1} & {\small{}0} & {\small{}0} & {\small{}0} & {\small{}0}\tabularnewline
{\small{}8} &  & {\small{}0} & {\small{}0} & {\small{}0} & {\small{}1} & {\small{}0} & {\small{}0} & {\small{}0} & {\small{}0} & {\small{}0} &  & {\small{}27} &  & {\small{}0} & {\small{}1} & {\small{}0} & {\small{}0} & {\small{}0} & {\small{}0} & {\small{}0} & {\small{}0} & {\small{}0}\tabularnewline
{\small{}9} &  & {\small{}0} & {\small{}1} & {\small{}0} & {\small{}0} & {\small{}0} & {\small{}0} & {\small{}0} & {\small{}0} & {\small{}0} &  & {\small{}28} &  & {\small{}0} & {\small{}0} & {\small{}0} & {\small{}0} & {\small{}0} & {\small{}1} & {\small{}1} & {\small{}0} & {\small{}0}\tabularnewline
{\small{}10} &  & {\small{}0} & {\small{}1} & {\small{}0} & {\small{}0} & {\small{}0} & {\small{}0} & {\small{}0} & {\small{}0} & {\small{}0} &  & {\small{}29} &  & {\small{}0} & {\small{}1} & {\small{}0} & {\small{}0} & {\small{}0} & {\small{}0} & {\small{}0} & {\small{}0} & {\small{}0}\tabularnewline
{\small{}11} &  & {\small{}1} & {\small{}0} & {\small{}0} & {\small{}0} & {\small{}0} & {\small{}1} & {\small{}0} & {\small{}0} & {\small{}0} &  & {\small{}30} &  & {\small{}0} & {\small{}0} & {\small{}0} & {\small{}1} & {\small{}0} & {\small{}0} & {\small{}0} & {\small{}0} & {\small{}1}\tabularnewline
{\small{}12} &  & {\small{}0} & {\small{}0} & {\small{}1} & {\small{}1} & {\small{}0} & {\small{}0} & {\small{}0} & {\small{}0} & {\small{}0} &  & {\small{}31} &  & {\small{}0} & {\small{}0} & {\small{}0} & {\small{}0} & {\small{}0} & {\small{}1} & {\small{}0} & {\small{}0} & {\small{}0}\tabularnewline
{\small{}13} &  & {\small{}0} & {\small{}0} & {\small{}0} & {\small{}0} & {\small{}0} & {\small{}0} & {\small{}0} & {\small{}1} & {\small{}0} &  & {\small{}32} &  & {\small{}1} & {\small{}0} & {\small{}0} & {\small{}0} & {\small{}0} & {\small{}1} & {\small{}0} & {\small{}0} & {\small{}0}\tabularnewline
{\small{}14} &  & {\small{}1} & {\small{}0} & {\small{}0} & {\small{}1} & {\small{}0} & {\small{}0} & {\small{}0} & {\small{}0} & {\small{}0} &  & {\small{}33} &  & {\small{}1} & {\small{}0} & {\small{}1} & {\small{}0} & {\small{}0} & {\small{}0} & {\small{}0} & {\small{}0} & {\small{}0}\tabularnewline
{\small{}15} &  & {\small{}0} & {\small{}0} & {\small{}0} & {\small{}0} & {\small{}0} & {\small{}1} & {\small{}0} & {\small{}0} & {\small{}0} &  & {\small{}34} &  & {\small{}0} & {\small{}0} & {\small{}0} & {\small{}0} & {\small{}1} & {\small{}0} & {\small{}0} & {\small{}0} & {\small{}0}\tabularnewline
{\small{}16} &  & {\small{}0} & {\small{}0} & {\small{}0} & {\small{}0} & {\small{}0} & {\small{}1} & {\small{}0} & {\small{}0} & {\small{}0} &  & {\small{}35} &  & {\small{}0} & {\small{}0} & {\small{}0} & {\small{}0} & {\small{}1} & {\small{}0} & {\small{}0} & {\small{}0} & {\small{}0}\tabularnewline
{\small{}17} &  & {\small{}0} & {\small{}0} & {\small{}0} & {\small{}1} & {\small{}0} & {\small{}0} & {\small{}0} & {\small{}1} & {\small{}0} &  & {\small{}36} &  & {\small{}0} & {\small{}0} & {\small{}1} & {\small{}1} & {\small{}0} & {\small{}0} & {\small{}0} & {\small{}0} & {\small{}0}\tabularnewline
{\small{}18} &  & {\small{}0} & {\small{}0} & {\small{}0} & {\small{}0} & {\small{}1} & {\small{}0} & {\small{}0} & {\small{}0} & {\small{}0} &  & {\small{}37} &  & {\small{}0} & {\small{}0} & {\small{}0} & {\small{}0} & {\small{}0} & {\small{}0} & {\small{}0} & {\small{}0} & {\small{}1}\tabularnewline
{\small{}19} &  & {\small{}0} & {\small{}1} & {\small{}0} & {\small{}0} & {\small{}0} & {\small{}0} & {\small{}0} & {\small{}0} & {\small{}0} &  &  &  &  &  &  &  &  &  &  &  & \tabularnewline
\hline 
\end{tabular*}\selectlanguage{english}%
\end{table}

\selectlanguage{american}%
\textbf{\vspace{-20pt}} 

The first step is to generate correlated attributes. Let $\boldsymbol{\theta}$
be the $N$ by $K$ underlying probability matrix of $\boldsymbol{\alpha}$,
and let column $k$ of $\boldsymbol{\theta}$ be a vector $\boldsymbol{\theta}_{k}$,
$k=1,\ldots,K$. That is, $\boldsymbol{\theta}=(\boldsymbol{\theta}{}_{1},\ldots,\boldsymbol{\theta}_{K})$.
A copula is used to generate intercorrelated $\boldsymbol{\theta}$
(see Ross, 2013). The correlation coefficient for each pair of columns
in $\boldsymbol{\theta}$ takes a constant value $\rho$ , and the
correlation matrix $\boldsymbol{\varSigma}$ is expressed as
\[
\boldsymbol{\varSigma}=\left[\begin{array}{ccc}
1 &  & \rho\\
 & \ddots\\
\rho &  & 1
\end{array}\right],
\]
where the off-diagonal entries are $\rho$. Each entry in $\boldsymbol{\varSigma}$
corresponds to the correlation coefficient between two columns in
$\boldsymbol{\theta}$. Symmetric with all the eigenvalues positive,
$\boldsymbol{\varSigma}$ is a real symmetric positive-definite matrix
that can be decomposed as $\boldsymbol{\varSigma}=\boldsymbol{\mathcal{\mathrm{\mathcal{\nu}}}}^{\mathrm{T}}\boldsymbol{\mathcal{\mathrm{\mathcal{\nu}}}}$
using Choleski decomposition, where $\mathcal{\boldsymbol{\mathcal{\mathrm{\mathcal{\nu}}}}}$
is an upper triangular matrix. 

After $\boldsymbol{\mathcal{\nu}}$ is derived, create an $I\times K$
matrix $\boldsymbol{\tau}$, in which each entry is generated from
$\mathit{N}(0,1)$. $\boldsymbol{\mathcal{\mathrm{\mathcal{\tau}}}}$
is then transformed to $\boldsymbol{\gamma}$ by using $\boldsymbol{\gamma}=\boldsymbol{\mathcal{\tau\boldsymbol{\nu}}}$,
so that $\boldsymbol{\mathcal{\mathrm{\gamma}}}$ and $\boldsymbol{\mathrm{\varSigma}}$
will have the same correlation structure. Set $\Phi(\boldsymbol{\mathcal{\mathrm{\gamma}}})=\boldsymbol{\theta}$,
where $\Phi(\cdot)$ is the cumulative standard normal distribution
function. $\boldsymbol{\alpha}$ is determined by
\[
\alpha_{ik}=\begin{cases}
1 & \textrm{if }\theta_{ik}\geq\Phi^{-1}(\frac{k}{K+1})\\
0 & \textrm{otherwise}
\end{cases},
\]
where $k=1,2,\ldots,K$ (see Chiu, Douglas, \& Li, 2009). Note that
the above method can also be used to generate correlated attributes
for the DINA and NIDA models.

The next step is to draw $\boldsymbol{\pi}^{*}$ and $\boldsymbol{r}^{*}$.
Set $g_{jk}$ and $s_{jk}$ to 0.2, and $\boldsymbol{\pi}^{*}$ and
$\boldsymbol{r}^{*}$ are obtained respectively from equations (1)
and (2). Probability of an examinee correctly answer an item is calculated
using equation (3), thus forming a matrix $\boldsymbol{y}=(y_{nj})_{N\times J}$.
The data is then generated using inverse transform sampling for two
points $0$ and $1$. Another matrix $\boldsymbol{\xi}=(\varepsilon_{nj})_{N\times J}$
is created where each element is generated from $\mathit{Uniform}(0,1)$,
and then $\boldsymbol{\xi}$ is compared with $\boldsymbol{y}$. If
the element in $\boldsymbol{\xi}$ is greater than the corresponding
element in $\boldsymbol{y}$ , set $y_{nj}$ to 0; if otherwise, then
set $y_{nj}$ to 1. The simulated data $\boldsymbol{y}$ is thus generated.

For $M$ simulated data sets, let $\hat{\boldsymbol{\alpha}}^{(m)}=(\hat{\alpha}_{nk}^{(m)})_{N\times K}$
$(m=1,\ldots,M)$ be the estimated Q-matrix from $m^{th}$ data set,
and let $\boldsymbol{\alpha}=(\alpha_{nk})_{N\times K}$ represents
the true $\boldsymbol{\alpha}$. To measure how well each method recovers
the true $\boldsymbol{\alpha}$, the measure of accuracy $\Delta_{\alpha}$,
confined between $0$ and $1$, is defined as 
\[
\Delta_{\alpha}=\frac{1}{M}\sum\limits _{m=1}^{M}\biggl(1-\frac{\biggl|\left[\hat{\boldsymbol{\alpha}}^{(m)}\right]-\boldsymbol{\alpha}\biggr|}{NK}\biggr),\:m=1,2,\ldots,M,
\]
where the $\left[\cdot\right]$ returns the value rounded to the nearest
integer and $|\cdot|$ is the absolute value. 

\noindent \textbf{Q-matrix in Simulation}

\noindent The Q-matrix (Table 1) for simulation I is obtained from
de la Torre (2008). 30 items that measure 5 attributes comprise this
artificial Q-matrix, which is constructed in a way that each attribute
appears alone, in a pair, or in triple the same number of times as
other attributes. This balanced Q-matrix, with each attribute being
measured by 12 items, appears to have a clear pattern that implies
main effects from items 1 to 10, two-way interactions from items 11
to 20 and three-way interactions from items 21 to 30. This Q-matrix
is complete, containing at least one item devoted solely to each attribute
(Chen, Liu, Xu, \& Ying, 2015). 

The Q-matrix (Table 2) for simulation II is acquired from Jang (2009),
which discusses second language speakers' reading comprehension. This
complex Q-matrix is imbalanced and incomplete, consisting of 37 items
that assess 9 attributes. For both simulations, examinees in groups
of 500, 1000 and 2000 are simulated with the correlation between each
pair of attributes set to 0.1, 0.3 and 0.5 for simplicity, as in Feng,
Habing, \& Huebner (2014). 20 data sets are simulated for each concoction.
Corresponding R codes are run 7000 iterations after 2000 burn-in periods. 

\noindent \textbf{Results}

\noindent The $\delta$ is set to 0.052 in step 4, so that the acceptance
rate is around 35\%. The Raftery and Lewis diagnostic (Raftery \&
Lewis, 1992) from the CODA R package (Plummer et al., 2006) suggests
that $\boldsymbol{\pi}^{*}$ and $\boldsymbol{r}^{*}$ estimates are
converged. Table 3 presents the results from simulations I and II.
For the complete and balanced Q-matrix in simulation I, the measure
of accuracy $\Delta_{\alpha}$ ranges from 0.919 to 0.941. For the
incomplete and imbalanced Q-matrix in simulation II, the $\Delta_{\alpha}$
is less than 0.9 but above 0.8, ranging from 0.822 to 0.843. 

It should be noted that using the independent model for simulation
I with sample size 2000 and correlation 0.5, we notice that the average
$\Delta_{\alpha}$ of 20 data sets drops to 0.835, indicating that
using the saturate model for correlated attributes is indeed improving
the accuracy of attribute estimates.
\begin{table}
\caption{Simulation Studies}

\centering%
\begin{tabular}{ccccccccccc}
\hline 
\noalign{\vskip\doublerulesep}
\multicolumn{5}{c}{{\small{}Simulation I}} &  & \multicolumn{5}{c}{{\small{}Simulation II}}\tabularnewline[\doublerulesep]
\cline{1-5} \cline{7-11} 
\noalign{\vskip\doublerulesep}
\noalign{\vskip\doublerulesep}
 &  & \multicolumn{3}{c}{{\small{}Correlation }} &  &  &  & \multicolumn{3}{c}{{\small{}Correlation }}\tabularnewline[\doublerulesep]
\cline{3-5} \cline{9-11} 
\noalign{\vskip\doublerulesep}
\noalign{\vskip\doublerulesep}
{\small{}Size} &  & {\small{}0.1} & {\small{}0.3} & {\small{}0.5} &  & {\small{}Size} &  & {\small{}0.1} & {\small{}0.3} & {\small{}0.5}\tabularnewline[\doublerulesep]
\cline{1-1} \cline{3-5} \cline{7-7} \cline{9-11} 
\noalign{\vskip\doublerulesep}
\noalign{\vskip\doublerulesep}
{\small{}500} &  & {\small{}0.919} & {\small{}0.925} & {\small{}0.928} &  & {\small{}500} &  & {\small{}0.822} & {\small{}0.829} & {\small{}0.834}\tabularnewline[\doublerulesep]
\cline{1-1} \cline{3-5} \cline{7-7} \cline{9-11} 
\noalign{\vskip\doublerulesep}
\noalign{\vskip\doublerulesep}
{\small{}1000} &  & {\small{}0.922} & {\small{}0.929} & {\small{}0.936} &  & {\small{}1000} &  & {\small{}0.829} & {\small{}0.832} & {\small{}0.837}\tabularnewline[\doublerulesep]
\cline{1-1} \cline{3-5} \cline{7-7} \cline{9-11} 
\noalign{\vskip\doublerulesep}
\noalign{\vskip\doublerulesep}
{\small{}2000} &  & {\small{}0.926} & {\small{}0.931} & {\small{}0.941} &  & {\small{}2000} &  & {\small{}0.835} & {\small{}0.839} & {\small{}0.843}\tabularnewline[\doublerulesep]
\hline 
\noalign{\vskip\doublerulesep}
\end{tabular}
\end{table}
\noindent \begin{center}
\textbf{\Large{}Empirical Study}
\par\end{center}{\Large \par}

\noindent Obtained from the CDM R package, the data consists of responses
of 2922 examinees to 28 multiple choice items that measure 3 attributes{\small{}
}(morphosyntactic, cohensive, lexical) in the grammar section of the
Examination for the Certificate of Proficiency in English (ECPE).
A standardized English as a foreign language examination, the ECPE
is recognized in several countries as official proof of advanced proficiency
in English (ECPE, 2015). 

The CDM R package is also used to compare with the results from the
MCMC algorithm. Specifically, the function with arguments \textit{gdina(data,
q.matrix, maxit=1000, rule=\textquotedbl{}RRUM\textquotedbl{})} is
applied. Note that the empirical Q-matrix (Table 5) is complete but
imbalanced.

Table 4 shows the classification rate of each attribute pattern. Table
5 exhibits parameter estimates from the MCMC algorithm and the CDM
R package. Applying the marginal maximum likelihood estimation, the
CDM R package is implemented using the EM algorithm. As can be seen
in Table 5, parameter estimates from the two methods do not deviate
much.

\begin{table}
\caption{Classification Rate}
\begin{centering}
\vspace{5pt}
\par\end{centering}
\begin{centering}
\centering%
\begin{tabular}{cccccccccc}
\hline 
 &  & \multicolumn{8}{c}{{\small{}Attribute Pattern}}\tabularnewline
\cline{3-10} 
{\small{}Method} &  & {\small{}(0,0,0)} & {\small{}(0,0,1)} & {\small{}(0,1,0)} & {\small{}(0,1,1)} & {\small{}(1,0,0)} & {\small{}(1,0,1)} & {\small{}(1,1,0)} & {\small{}(1,1,1)}\tabularnewline
\cline{1-1} \cline{3-10} 
\noalign{\vskip\doublerulesep}
{\small{}MCMC} &  & {\small{}0.309} & {\small{}0.120} & {\small{}0.006} & {\small{}0.186} & {\small{}0.004} & {\small{}0.007} & {\small{}0.002} & {\small{}0.369}\tabularnewline
\cline{1-1} \cline{3-10} 
\noalign{\vskip\doublerulesep}
{\small{}CDM R} &  & {\small{}0.294} & {\small{}0.124} & {\small{}0.020} & {\small{}0.181} & {\small{}0.010} & {\small{}0.013} & {\small{}0.008} & {\small{}0.353}\tabularnewline[\doublerulesep]
\hline 
\end{tabular}
\par\end{centering}
\noindent \begin{raggedright}
{\small{}\vspace{-4pt}}
\par\end{raggedright}{\small \par}
\noindent \raggedright{}\textit{\small{}\hspace{30pt}}\textit{\footnotesize{}Note}{\footnotesize{}.
CDM R stands for the CDM R package}{\footnotesize \par}
\end{table}

{\small{}}
\begin{table}
{\small{}\caption{Empirical Study}
\vspace{-8pt}}{\small \par}

{\small{}\centering}%
\begin{tabular}[t]{>{\centering}p{20pt}cccccc>{\centering}p{0.5pt}ccccccccc}
\hline 
\noalign{\vskip\doublerulesep}
 &  & \multicolumn{3}{c}{{\footnotesize{}Q-matrix}} &  & \multicolumn{5}{c}{{\footnotesize{}MCMC}} &  & \multicolumn{5}{c}{{\footnotesize{}CDM R}}\tabularnewline[\doublerulesep]
\cline{3-5} \cline{7-11} \cline{13-17} 
\noalign{\vskip\doublerulesep}
 &  & \multirow{2}{*}{{\footnotesize{}Mor}} & \multirow{2}{*}{{\footnotesize{}Coh}} & \multirow{2}{*}{{\footnotesize{}Lex}} &  & \multirow{2}{*}{{\footnotesize{}$\pi^{*}$}} &  &  & {\footnotesize{}$r^{*}$} &  &  & \multirow{2}{*}{{\footnotesize{}$\pi^{*}$}} &  & \multicolumn{3}{c}{{\footnotesize{}$r^{*}$}}\tabularnewline
\cline{9-11} \cline{15-17} 
{\footnotesize{}Item} &  &  &  &  &  &  &  & {\footnotesize{}Mor} & {\footnotesize{}Coh} & {\footnotesize{}Lex} &  &  &  & {\footnotesize{}Mor} & {\footnotesize{}Coh} & {\footnotesize{}Lex}\tabularnewline
\cline{1-1} \cline{3-5} \cline{7-11} \cline{13-17} 
{\footnotesize{}E1} &  & {\footnotesize{}1} & {\footnotesize{}1} & {\footnotesize{}0} &  & {\footnotesize{}0.926 } &  & {\footnotesize{}0.876 } & {\footnotesize{}0.853 } &  &  & {\footnotesize{}0.928 } &  & {\footnotesize{}0.875 } & {\footnotesize{}0.851 } & \tabularnewline
{\footnotesize{}E2} &  & {\footnotesize{}0} & {\footnotesize{}1} & {\footnotesize{}0} &  & {\footnotesize{}0.906 } &  &  & {\footnotesize{}0.813 } & {\footnotesize{} } &  & {\footnotesize{}0.905 } &  &  & {\footnotesize{}0.812 } & \tabularnewline
{\footnotesize{}E3} &  & {\footnotesize{}1} & {\footnotesize{}0} & {\footnotesize{}1} &  & {\footnotesize{}0.780 } &  & {\footnotesize{}0.636 } &  & {\footnotesize{}0.833 } &  & {\footnotesize{}0.784 } &  & {\footnotesize{}0.640 } &  & {\footnotesize{}0.824 }\tabularnewline
{\footnotesize{}E4} &  & {\footnotesize{}0} & {\footnotesize{}0} & {\footnotesize{}1} &  & {\footnotesize{}0.824 } &  &  &  & {\footnotesize{}0.564 } &  & {\footnotesize{}0.825 } &  &  &  & {\footnotesize{}0.562 }\tabularnewline
{\footnotesize{}E5} &  & {\footnotesize{}0} & {\footnotesize{}0} & {\footnotesize{}1} &  & {\footnotesize{}0.956 } &  &  &  & {\footnotesize{}0.779 } &  & {\footnotesize{}0.957 } &  &  &  & {\footnotesize{}0.779 }\tabularnewline
{\footnotesize{}E6} &  & {\footnotesize{}0} & {\footnotesize{}0} & {\footnotesize{}1} &  & {\footnotesize{}0.926 } &  &  &  & {\footnotesize{}0.760 } &  & {\footnotesize{}0.927 } &  &  &  & {\footnotesize{}0.760 }\tabularnewline
{\footnotesize{}E7} &  & {\footnotesize{}1} & {\footnotesize{}0} & {\footnotesize{}1} &  & {\footnotesize{}0.940 } &  & {\footnotesize{}0.737 } &  & {\footnotesize{}0.704 } &  & {\footnotesize{}0.943 } &  & {\footnotesize{}0.738 } &  & {\footnotesize{}0.704 }\tabularnewline
{\footnotesize{}E8} &  & {\footnotesize{}0} & {\footnotesize{}1} & {\footnotesize{}0} &  & {\footnotesize{}0.966 } &  &  & {\footnotesize{}0.841 } &  &  & {\footnotesize{}0.966 } &  &  & {\footnotesize{}0.840 } & \tabularnewline
{\footnotesize{}E9} &  & {\footnotesize{}0} & {\footnotesize{}0} & {\footnotesize{}1} &  & {\footnotesize{}0.787 } &  &  &  & {\footnotesize{}0.673 } &  & {\footnotesize{}0.788 } &  &  &  & {\footnotesize{}0.672 }\tabularnewline
{\footnotesize{}E10} &  & {\footnotesize{}1} & {\footnotesize{}0} & {\footnotesize{}0} &  & {\footnotesize{}0.888 } &  & {\footnotesize{}0.574 } &  &  &  & {\footnotesize{}0.892 } &  & {\footnotesize{}0.575 } &  & \tabularnewline
{\footnotesize{}E11} &  & {\footnotesize{}1} & {\footnotesize{}0} & {\footnotesize{}1} &  & {\footnotesize{}0.924 } &  & {\footnotesize{}0.763 } &  & {\footnotesize{}0.701 } &  & {\footnotesize{}0.925 } &  & {\footnotesize{}0.769 } &  & {\footnotesize{}0.695 }\tabularnewline
{\footnotesize{}E12} &  & {\footnotesize{}1} & {\footnotesize{}0} & {\footnotesize{}1} &  & {\footnotesize{}0.728 } &  & {\footnotesize{}0.522 } &  & {\footnotesize{}0.371 } &  & {\footnotesize{}0.733 } &  & {\footnotesize{}0.527 } &  & {\footnotesize{}0.362 }\tabularnewline
{\footnotesize{}E13} &  & {\footnotesize{}1} & {\footnotesize{}0} & {\footnotesize{}0} &  & {\footnotesize{}0.905 } &  & {\footnotesize{}0.726 } &  &  &  & {\footnotesize{}0.907 } &  & {\footnotesize{}0.727 } &  & \tabularnewline
{\footnotesize{}E14} &  & {\footnotesize{}1} & {\footnotesize{}0} & {\footnotesize{}0} &  & {\footnotesize{}0.821 } &  & {\footnotesize{}0.660 } &  &  &  & {\footnotesize{}0.826 } &  & {\footnotesize{}0.658 } &  & \tabularnewline
{\footnotesize{}E15} &  & {\footnotesize{}0} & {\footnotesize{}0} & {\footnotesize{}1} &  & {\footnotesize{}0.957 } &  &  &  & {\footnotesize{}0.761 } &  & {\footnotesize{}0.958 } &  &  &  & {\footnotesize{}0.761 }\tabularnewline
{\footnotesize{}E16} &  & {\footnotesize{}1} & {\footnotesize{}0} & {\footnotesize{}1} &  & {\footnotesize{}0.906 } &  & {\footnotesize{}0.751 } &  & {\footnotesize{}0.715 } &  & {\footnotesize{}0.909 } &  & {\footnotesize{}0.753 } &  & {\footnotesize{}0.714 }\tabularnewline
{\footnotesize{}E17} &  & {\footnotesize{}0} & {\footnotesize{}1} & {\footnotesize{}1} &  & {\footnotesize{}0.943 } &  &  & {\footnotesize{}0.916 } & {\footnotesize{}0.923 } &  & {\footnotesize{}0.943 } &  &  & {\footnotesize{}0.919 } & {\footnotesize{}0.920 }\tabularnewline
{\footnotesize{}E18} &  & {\footnotesize{}0} & {\footnotesize{}0} & {\footnotesize{}1} &  & {\footnotesize{}0.910 } &  &  &  & {\footnotesize{}0.785 } &  & {\footnotesize{}0.910 } &  &  &  & {\footnotesize{}0.785 }\tabularnewline
{\footnotesize{}E19} &  & {\footnotesize{}0} & {\footnotesize{}0} & {\footnotesize{}1} &  & {\footnotesize{}0.838 } &  &  &  & {\footnotesize{}0.537 } &  & {\footnotesize{}0.839 } &  &  &  & {\footnotesize{}0.538 }\tabularnewline
{\footnotesize{}E20} &  & {\footnotesize{}1} & {\footnotesize{}0} & {\footnotesize{}1} &  & {\footnotesize{}0.754 } &  & {\footnotesize{}0.500 } &  & {\footnotesize{}0.516 } &  & {\footnotesize{}0.759 } &  & {\footnotesize{}0.501 } &  & {\footnotesize{}0.511 }\tabularnewline
{\footnotesize{}E21} &  & {\footnotesize{}1} & {\footnotesize{}0} & {\footnotesize{}1} &  & {\footnotesize{}0.917 } &  & {\footnotesize{}0.849 } &  & {\footnotesize{}0.705 } &  & {\footnotesize{}0.917 } &  & {\footnotesize{}0.854 } &  & {\footnotesize{}0.699 }\tabularnewline
{\footnotesize{}E22} &  & {\footnotesize{}0} & {\footnotesize{}0} & {\footnotesize{}1} &  & {\footnotesize{}0.796 } &  &  &  & {\footnotesize{}0.371 } &  & {\footnotesize{}0.797 } &  &  &  & {\footnotesize{}0.370 }\tabularnewline
{\footnotesize{}E23} &  & {\footnotesize{}0} & {\footnotesize{}1} & {\footnotesize{}0} &  & {\footnotesize{}0.936 } &  &  & {\footnotesize{}0.704 } &  &  & {\footnotesize{}0.936 } &  &  & {\footnotesize{}0.699 } & \tabularnewline
{\footnotesize{}E24} &  & {\footnotesize{}0} & {\footnotesize{}1} & {\footnotesize{}0} &  & {\footnotesize{}0.698 } &  &  & {\footnotesize{}0.479 } &  &  & {\footnotesize{}0.696 } &  &  & {\footnotesize{}0.474 } & \tabularnewline
{\footnotesize{}E25} &  & {\footnotesize{}1} & {\footnotesize{}0} & {\footnotesize{}0} &  & {\footnotesize{}0.771 } &  & {\footnotesize{}0.676 } &  &  &  & {\footnotesize{}0.775 } &  & {\footnotesize{}0.674 } &  & \tabularnewline
{\footnotesize{}E26} &  & {\footnotesize{}0} & {\footnotesize{}0} & {\footnotesize{}1} &  & {\footnotesize{}0.782 } &  &  &  & {\footnotesize{}0.691 } &  & {\footnotesize{}0.783 } &  &  &  & {\footnotesize{}0.690 }\tabularnewline
{\footnotesize{}E27} &  & {\footnotesize{}1} & {\footnotesize{}0} & {\footnotesize{}0} &  & {\footnotesize{}0.689 } &  & {\footnotesize{}0.422 } &  &  &  & {\footnotesize{}0.695 } &  & {\footnotesize{}0.421 } &  & \tabularnewline
{\footnotesize{}E28} &  & {\footnotesize{}0} & {\footnotesize{}0} & {\footnotesize{}1} &  & {\footnotesize{}0.909 } &  &  &  & {\footnotesize{}0.701 } &  & {\footnotesize{}0.910 } &  &  &  & {\footnotesize{}0.701 }\tabularnewline
\hline 
\end{tabular}{\small \par}

{\small{}\vspace{1.5pt}}\textit{\footnotesize{}Note}{\footnotesize{}.
Mor = morphosyntactic; Coh = cohensive; Lex = lexical}{\footnotesize \par}
\end{table}
{\small{} }{\small \par}
\noindent \begin{center}
\textbf{\Large{}Discussion}
\par\end{center}{\Large \par}

\noindent The current research proposes an MCMC algorithm for estimating
parameters of the RRUM in a Bayesian framework. The algorithm is summarized
as follows. Using the binary decimal conversion, possible attribute
patterns are transformed to a Multinomial distribution (step 1). Along
with the likelihood of an attribute pattern, a Dirichlet distribution
is used as the prior to sample from the posterior. The Dirichlet distribution
is constructed using Gamma distributions (step 2), and attributes
of examinees are updated using the inverse transform sampling (step
3). Sequentially, $\boldsymbol{r}^{*}$ and $\boldsymbol{\pi}^{*}$
are generated using the Metropolis within Gibbs sampler (step 4).
Of note is that steps 1 to 3 can also be used in estimating correlated
attributes in the DINA and NIDA models.

Like most of the studies, the first simulation uses a complete and
balanced Q-matrix. The measure of accuracy is on average 0.929. However
when the Q-matrix is incomplete and imbalanced as in the second simulation,
the measure of accuracy drops to an average of 0.833. A similar result
is also revealed using the EM algorithm in the CDM R package. Therefore,
one should be cautious when using a complex Q-matrix for the RRUM.

Another issue is the correlation between each pair of attributes.
As can be seen from Table 3, when the sample size increases, the measure
of accuracy increases as expected. However, when the correlation between
each pair of attributes is higher, the measure of accuracy is counterintuitively
lower. Chen, Liu, Xu, \& Ying (2015) also observes a similar phenomenon
in their Q-matrix research based on the DINA model. A heuristic explanation
according to Chen, Liu, Xu, \& Ying (2015) is that the simulated data
has more observations with the attribute pattern $(0,0,0,0,0)$ when
the correlation is higher. For a sample size of 1000 in simulation
I, there are around 90, 60 and 20 examinees having attribute pattern
$(0,0,0,0,0)$ for correlations 0.5, 0.3 and 0.1, respectively. Because
it is more difficult to identify $(0,0,0,0,0)$, the algorithm needs
more $(0,0,0,0,0)$ examinees to estimate accurately. Therefore when
the the correlation is higher, the the performance of the algorithm
is better.

The complete Q-matrix for the empirical study measures only 3 attributes
although imbalanced. The result is consistent with that from the CDM
R package and from Feng, Habing, \& Huebner (2014). It is suggested
that future research compare the estimated examinees' attribute patterns
with the estimate from other CDMs such as the popular DINA model.\newpage{}
\noindent \begin{center}
\textbf{\Large{}References}\vspace{-30pt}
\par\end{center}
\begin{description}
\begin{singlespace}
\item [{\textmd{Chen,}}] Y., Liu, J., Xu, G., \& Ying, Z. (2015). Statistical
analysis of Q-matrix based diagnostic classification models. \textit{Journal
of the American Statistical Association}, 110(510), 850-866.
\item [{\textmd{Chiu,}}] C.-Y, Douglas J., \& Li, X. (2009). Cluster analysis
for cognitive diagnosis: Theory and applications. \textit{Psychometrika},
74, 633-665.
\item [{\textmd{Chiu,}}] C.-Y., \& Köhn, H.-F. (2016). The reduced RUM
as a logit model: Parameterization and constraints.\textit{ Psychometrika},
81: 350.
\item [{\textmd{de}}] la Torre, J. (2008). An empirically-based method
of Q-matrix validation for the DINA model: Development and applications.
\textit{Journal of Educational Measurement}, 45, 343-362.
\item [{\textmd{DiBello,}}] L., Roussos, L. A., \& Stout, W. (2007). Review
of cognitively diagnostic assessment and a summary of psychometric
models. In C. V. Rao \& S. Sinharay (Eds.), \textit{Handbook of statistics}
(Vol. 26, Psychometrics, pp. 979-1027). Amsterdam, the Netherlands:
Elsevier.
\item [{\textmd{ECPE}}] (2015). ECPE 2015 Report (p. 1). The Examination
for the Certificate of Proficiency in English (ECPE). Retrieved from
http://cambridgemichigan.org/institutions/products-services/tests/proficiency-certification/ecpe/
\item [{\textmd{Feng,}}] Y., Habing, B. T., \& Huebner, A. (2014). Parameter
estimation of the Reduced RUM using the EM algorithm.\textit{ Applied
Psychological Measurement}, 38, 137\textendash 150.
\item [{\textmd{Hartz,}}] S. (2002). \textit{A Bayesian framework for the
Unified Model for assessing cognitive abilities: Blending theory with
practicality} (Doctoral dissertation). University of Illinois, Urbana-Champaign.
\item [{\textmd{Hartz,}}] S., Roussos, L., \& Stout, W. (2002). Skills
diagnosis: Theory and practice. Unpublished manuscript. University
of Illinois at Urbana Champaign.
\item [{\textmd{Henson,}}] R., \& Templin, J. (2007, April). \textit{Importance
of Q-matrix construction and its effects cognitive diagnosis model
results}. Paper presented at the annual meeting of the National Council
on Measurement in Education in Chicago, Illinois.
\item [{\textmd{Jang,}}] E. E. (2009). Cognitive diagnostic assessment
of L2 reading comprehension ability: Validity arguments for applying
Fusion Model to LanguEdge assessment.\textit{ Language Testing}, 26(1),
31\textendash 73.
\item [{\textmd{Jang,}}] E. E., Dunlop, M., Wagner, M., Kim, Y. H., \&
Gu, Z. (2013). Elementary school ELLs\textquoteright{} reading skill
profiles using cognitive diagnosis modeling: Roles of length of residence
and home language environment. \textit{Language Learning}, 63(3),
400\textendash 436.
\item [{\textmd{Junker,}}] B.W., \& Sijtsma, K. (2001). Cognitive assessment
models with few assumptions, and connections with nonparametric item
response theory. \textit{Applied Psychological Measurement}, 25, 258-272.
\item [{\textmd{Kim,}}] Y. H. (2011). Diagnosing EAP writing ability using
the reduced Reparameterized Unified Model. \textit{Language Testing},
28, 509\textendash 541.
\item [{\textmd{Kim,}}] A. Y. (2014). Exploring ways to provide diagnostic
feedback with an ESL placement test: Cognitive diagnostic assessment
of L2 reading ability. \textit{Language Testing}, 32(2): 227\textendash 258.
\item [{\textmd{Leighton,}}] J.P., \& Gierl, M.J. (Eds.). (2007). \textit{Cognitive
diagnostic assessment for education. Theory and applications}. Cambridge,
MA: Cambridge University Press.
\item [{\textmd{Li,}}] H., \& Suen, H. K. (2013). Constructing and validating
a Q-matrix for cognitive diagnostic analyses of a reading test. \textit{Educational
Assessment}, 18(1), 1\textendash 25.
\item [{\textmd{Maris,}}] E. (1999). Estimating multiple classification
latent class models. \textit{Psychometrika}, 64, 187\textendash 212.
\item [{\textmd{Plummer,}}] M., Best, N., Cowles, K., \& Vines, K. (2006).
CODA: Convergence Diagnosis and Output Analysis for MCMC, R News,
vol 6, 7-11
\item [{\textmd{R}}] Development Core Team. (2017). R: A language and environment
for statistical computing {[}Computer software{]}. Vienna, Austria:
R Foundation for Statistical Computing. Available from http://www.r-project.org.
\item [{\textmd{Raftery,}}] A.E. \& Lewis, S.M. (1992). One long run with
diagnostics: Implementation strategies for Markov chain Monte Carlo.
\textit{Statistical Science}, 7, 493-497.
\item [{\textmd{Richards,}}] B. (2008). Formative Assessment in Teacher
Education: The Development of a Diagnostic Language Test for Trainee
Teachers of German. \textit{British Journal of Educational Studies},
56(2), 184-204.
\item [{\textmd{Ross,}}] S. M. (2006), \textit{Simulation}. 4th ed., Academic
Press, San Diego.
\item [{\textmd{Roussos,}}] L. A., Templin, J. L., \& Henson, R. A. (2007).
Skills diagnosis using IRT-based latent class models. \textit{Journal
of Educational Measurement}, 44, 293-311.
\item [{\textmd{Tatsuoka,}}] K. K. (1983). Rule space: An approach for
dealing with misconceptions based on item response theory. \textit{Journal
of Educational Measurement}, 20, 345\textendash 354.
\item [{\textmd{Templin,}}] J. (2004). Estimation of the RUM without alpha
tilde: a general model for the proficiency space of examinee ability.
External Diagnostic Research Group Technical Report.
\item [{\textmd{Templin,}}] J., \& Douglas, J. (2004). Higher order RUM.
External Diagnostic Research Group Technical Report
\item [{\textmd{Templin,}}] J., Henson, R., Templin, S., \& Roussos, L.
(2004). Robustness of unidimensional hierarchical modeling of discrete
attribute association in cognitive diagnosis models. External Diagnostic
Research Group Technical Report.
\end{singlespace}
\end{description}

\end{document}